# Proposal for photoacoustic ultrasonic generator based on Tamm plasmon structures

Elizaveta I. Girshova[1], Alena P. Mikitchuk[2], Alexey V. Belonovski[3], Konstantin M. Morozov[1], Konstantin A. Ivanov[3], Galia Pozina[4], Konstantin V. Kozadaev[2], Anton Yu. Egorov[3], and Mikhail A. Kaliteevski[1,3]

[1] *St Petersburg Academic University, Khlopina Str. 8/3, 194021 St Petersburg, Russia*
[2] *Belarusian State University, Niezaliezhnasci Avenue 4, 220030 Minsk, Belarus*
[3] *ITMO University, Kronverksky Pr. 49, 197101 St. Petersburg, Russia*
[4] *Department of Physics, Chemistry and Biology, Linköping University, 58183 Linköping, Sweden*
*Corresponding author: galia.pozina@liu.se*

**The scheme of generation of ultrasound waves based on optically excited Tamm plasmon structures is proposed. It is shown that Tamm plasmon structures can provide total absorption of a laser pulse with arbitrary wavelength in a metallic layer providing the possibility of the use of an infrared semiconductor laser for the excitation of ultrasound waves. Laser pulse absorption, heat transfer and dynamical properties of the structure are modeled, and the optimal design of the structure is found. It is demonstrated that the Tamm plasmon-based photoacoustic generator can emit ultrasound waves in the frequency band up to 100 MHz with pre-defined frequency spectrum. Optical power to sound power conversion efficiency grows linearly with frequency of the laser modulation and excitation power.** © 2020 Optical Society of America

http://dx.doi.org/

Ultrasound waves are widely used for ultrasonic non-destructive evaluation, structure health monitoring, material characterization [1], orientation imaging microscopy [2] and acoustic microscopy [3, 4] and optoacoustic tomography [5]. Existing conventional approach for generation of ultrasound waves utilize piezoelectric transducers, which bring some drawbacks associated with poor survivability in harsh environment, bulky size and weight, complexity of setups, high cost and susceptibility to electromagnetic interference [1]. Additionally, piezoelectric transducers exhibit a spectrum centered at the resonance frequency [6]. In contrary, photoacoustic transducers are a very attractive alternative for generation of ultrasound, because they are based on the effect of expansion of the optical absorbing layer heated by laser pulses [7,8].

In photoacoustic transducers, the absorber is heated and cooled, leading to mechanical deformations, which cause cycles of pressure, or, in the other words, acoustic waves in ambient surrounding [9]. The operation principle provides the number of advantages: reliability, compactness, full galvanic isolation [10], wideband operation [11], and possibility of realization of transducer at the edge of the optical fiber [12], that, in turn, leads to the immunity to harsh environment and electromagnetic interference. At the same time, further miniaturization, ultrawideband realization as well as the increase of functional flexibility of the ultrasound generation is required. Such goals can be achieved by the application of plasmonic structures, especially the Tamm plasmon (TP) structures [13,14].

TP is the state of electromagnetic field localized at the interface between metal and the Bragg reflector (BR). TP structures provide a wide range of opportunities for electromagnetic mode engineering [15,16] and, in particular, for the control of the absorption of light [17]. In optoelectronic devices, absorption of light in metals is the enemy of high performance, but for photoacoustic generators, high performance of the device is based on controllable absorption of light.

In existing photoacoustic systems, the laser wavelength usually corresponds to green-blue visible range that limits the range of possible laser systems. The use of TP structures gives opportunity to provide a total absorption in the infrared band. In this case, GaAs-based semiconductor lasers emitting at a wavelength of 980 nm can be used. Such lasers combine affordability, high power, scalability, and possibility of a direct temporal modulation of the intensity for the frequencies up to few GHz [18,19]. Semiconductor lasers operating at the wavelength of 980 nm can produce an average power of 1 W focused on the 10 μm x 10 μm spot providing a flux density up to $10^6$ W/cm$^2$.

This letter aims to introduce the novel scheme of the photoacoustic generator based on the Tamm plasmon structure (PAG-TP) and the optimisation of its performance. In the proposed scheme the metallic layer of the TP structure is heated by a short laser pulse, expands and emits ultrasound waves, as shown in Fig. 1 Utilisation of the TP enhances the absorption of light in a predefined point of the structure and could improve the performance of the photoacoustic generator of ultrasound waves.

The structure consists of BR covered by a layer of metal and is illuminated from the BR side by the laser beam periodically modulated in time. The design of the TP structure ensures the total absorption of laser radiation at a desirable wavelength resulting to a periodic in time heating and cooling of the metallic layer and, therefore, its periodic expansion and shrinking leading to emission of sound.

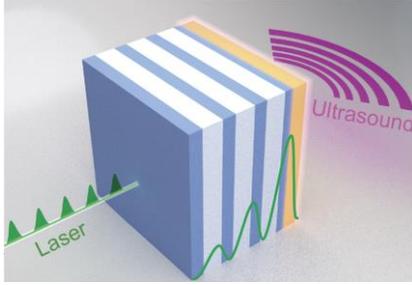

**Fig. 1.** Scheme of the PAG-TP structure and the principle of operation.

The design of the structure and the choice of the material should simultaneously provide: (i) total absorption of laser radiation at the desired wavelength; (ii) effective heating of the metallic layer by laser pulses; (iii) effective heat sink from the metallic layer in order to avoid overheating; (iv) maximal amplitude and desirable temporal shape of oscillation of the sample surface.

When the layer of the material with thermal expansion coefficient $\varepsilon$ absorbs the laser pulse with energy density G, the thickness of the layer is increased by the value $B$ defined by

$$B = \varepsilon G/(c\rho), \qquad (1)$$

where $c$ and $\rho$ are the specific heat capacity and density of the material, respectively. In the case of a harmonic oscillation of the surface with amplitude of oscillation $B$ and frequency $f$ the intensity of sound $J$ reads as [20]:

$$J = \frac{\rho_m v}{2}(2\pi f B)^2, \qquad (2)$$

where $\rho_m$ and $v$ are the density and speed of sound for the media, where ultrasound is emitted. Thus, the efficiency of conversion of laser radiation to ultrasound waves $\eta$ can be estimated as

$$\eta = \frac{J}{Gf} = 2\pi^2 \rho_m v \left(\frac{\varepsilon}{c\rho}\right)^2 Gf. \qquad (3)$$

As shown by equation 3, the key parameter defining the efficiency $\eta$ is the ratio $\varepsilon/(c\rho)$ for the metal forming a top layer of the structure. Usually, metals like gold, silver or copper are used to fabricate photoacoustic emitters but as can be seen from Table 1, lithium (Li) possesses maximal value of $\varepsilon/(c\rho)$ followed by lead (Pb) and magnesium (Mg). Magnesium possesses a larger thermal diffusivity coefficient than lithium or lead. Therefore magnesium (or its alloys) is the most suitable material for metallic layer in PAG-TP. For the materials forming the Bragg reflector, it is convenient to choose TiO$_2$ as the high refractive index material with large thermal diffusivity coefficient, and SiO$_2$ as the low refractive index materials.

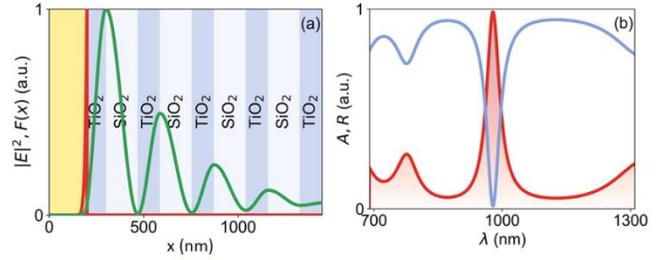

**Fig. 2.** (a)Scheme of the TP structure. Profile of $|E(x)|^2$ established in the TP structure under illumination by the laser with wavelength of 980 nm (green line) and the corresponding heating density (red line). (b) Absorption (red) and reflection (blue) spectra of the TP structure under illumination from BR side.

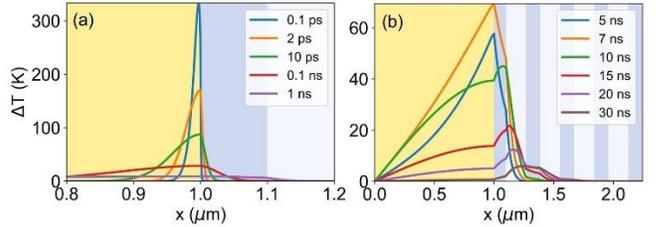

**Fig. 3.** Modification of the spatial profile of the temperature in the structure after heating by a single laser pulse with heating density profile shown in figure 2a, for infinitely short pulse with energy 1 mJ/cm$^2$ (a) and pulse of sinusoidal shape with duration 10 ns(b) and energy and 10 mJ/cm$^2$.

Table 1 shows the properties of some metals and dielectric materials [21-26], which can be used to construct PAG-TP. Figure 2(a) shows the detailed scheme of the TP structure. The structure consists of a magnesium film deposited on top of the Bragg reflector made of 4 pairs of SiO$_2$ and TiO$_2$ layers with a thickness of 168 nm and 118 nm, respectively. The thickness of the layer of TiO$_2$ adjacent to the magnesium layer is reduced to 100 nm in order to shift the TP to the center of the stopband of the Bragg reflector and provide better localization of light. The thickness of the magnesium layer is 1000 nm.

Spectra of absorption $A(\lambda)$ and reflection $R(\lambda)$ coefficients of the TP structure are shown in Fig. 2(b) [27]. Clearly, there is a sharp peak of absorption at the wavelength of 980 nm corresponding to the TP resonance, with the maximum value being close to unity. Note that when the light is incident on the interface of magnesium from the uniform material, the reflection coefficient at the wavelength of 980 nm is ~0.8. The green line in Fig. 2(a) shows the spatial profile of $|E(x)|^2$, where $E(x)$ is electric field established within the structure under illumination by the light with wavelength of 980 nm. It can be seen that the electric field is localized near the interface of the metal and the penetration depth is $\xi \approx 15$ nm. Penetration of laser radiation into metal and its absorption lead to the generation of the heat with a density F(x,t) described by

$$F(x,t) = \frac{A(\lambda)I(t)\alpha(x,\lambda)|E(x)|^2}{\int \alpha(x)|E(x)|^2 dx}, \qquad (4)$$

**Table 1.** Specific heat capacity $c$, thermal conductivity $\kappa$, density $\rho$, thermal diffusivity D, thermal expansion coefficient $\varepsilon$, and quantity $\frac{\varepsilon}{c\rho}$ of the materials used to construct TP structure.

|  | $c$ | $\kappa$ | $\rho$ | $c\rho$ | D | $\varepsilon$ | $\frac{\varepsilon}{c\rho}$ |
|---|---|---|---|---|---|---|---|
|  | J/(kg K) | W/(m K) | $10^3$ (kg/m³) | $10^6$ J/(m³K) | $10^{-6}$ m²/c | $10^{-6}$ K⁻¹ | $10^{-12}$ m³/J |
| Au | 128 | 317 | 19.3 | 2.48 | 127 | 14.2 | 5.7 |
| Ag | 235 | 235 | 10.4 | 2.45 | 96.15 | 19.5 | 7.9 |
| Al | 897 | 236 | 2.71 | 2.43 | 97 | 22 | 9.4 |
| Mg | 103 | 156 | 1.73 | 1.78 | 88 | 25 | 15 |
| Cu | 385 | 401 | 8.92 | 3.44 | 116 | 16.6 | 4.0 |
| Pb | 130 | 35.3 | 11.3 | 1.45 | 30 | 28 | 19 |
| SiO₂ | 772 | 1.38 | 2.2 | 1.698 | 0.812 | 0.5 | 0.29 |
| TiO₂ | 711. | 12.6 | 4.26 | 3.032 | 4.15 | 9.19 | 3.03 |
| Al₂O₃ | 850 | 30 | 3.99 | 3.3915 | 8.84 | 8.1 | 2.38 |

where $I(t, \lambda)$ is an intensity of incident light and $\alpha(x, \lambda)$ is the light absorption coefficient for the wavelength $\lambda$. The profile of the $F(x, t)$ is shown in Fig. 2(a) by the red line. Temporal variation of the temperature profile within the structure can be found by solving inhomogeneous heat equation:

$$c\rho \frac{dT}{dt} = \frac{d}{dx}\left(\kappa \frac{dT}{dx}\right) + F(x,t) \quad , \tag{5}$$

where $c$ is a specific heat capacity, $\kappa$ is the thermal conductivity, $\rho$ is the density. Sometimes it is suitable to use homogeneous heat equation in the form

$$\frac{dT}{dt} = D \frac{d^2T}{dx^2} \quad , \tag{6}$$

where $D = \kappa/(c\rho)$ is the thermal diffusivity. Performance of the PAG-TP is defined by the dynamic of the temperature distribution, and for a quantitative understanding of the temperature relaxation processes, we first consider heating of the structure by an infinitely short laser pulse with energy density $G$. Since the materials of Bragg reflector adjacent to the metal does not absorb light, the short single laser pulse is absorbed only in metal layer, leading to the formation of temperature profile defined by product $\alpha(x)|E(x)|^2$ shown in Fig. 2(a). In the subsequent modeling we assume that PAG-TP is confined from the both sides by the heat sinks supporting room temperature. After the end of the heating laser pulse the structure starts to cool. We can estimate the characteristic time of the temperature relaxation in the structure using well known relation

$$\langle x^2 \rangle = 2Dt \tag{7}$$

The thermal diffusivity of the metals is more than one order of magnitude larger than for SiO₂ and TiO₂. Thus, according to equation 4, the relaxation of the temperature within the metal will occur much faster than in the BR. At the initial stage of cooling, the temperature will fall by factor of two in the time $\tau_1 = \frac{\xi^2}{2D_M}$, which is 3 ps for the structure under study. Then, the temporal dependence of the maximal temperature increase, $\Delta T$, in the sample is governed by the fundamental solution for the thermal conductivity equation, requiring that

$$\Delta T \sim \sqrt{1/t} \tag{8}$$

and then the leveling of the temperature within the metal layer occurs approximately within a characteristic time $\tau_2 = \frac{d_M^2}{2D_M}$. For the thickness of metal $d_M = 1\ \mu$m, the spreading of heat within the metallic layer occurs within ~7 ns. Figure 3(a) shows the temporal evolution of the temperature profile in the structure obtained by numerical solving equation 6 [28] after heating by an infinitely sort laser pulse. The initial distribution of temperature profile repeats the profile of heat density $F(x)$. Then the temperature is reducing according to equation 8, and the profile is spreading; $\Delta T$ reduces by a factor of two during 2 ps, while leveling of the temperature in the metallic layer occurs within 1 ns. Note that most of the heat goes into the metal, but only a small part of the BR (with thickness of ~100 nm) adjacent to the metal has a temperature comparable to those of metal. Figure 3(b) shows the evolution of the temperature increase $\Delta T$ under heating of the structure with a single laser pulse of sinusoidal shape with a duration of 10 ns obtained numerical solving eq. (5). During the first half of the heating pulse, when the density of the pulse increases, the temperature profile has a maximum at the edge of the metallic layers. On the other hand, the temperature of the BR layers adjacent to the metal, increases. After the end of heating pulse, at 10 ns, the maximum of temperature falls within the BR. Due to a small thermal diffusivity of the BR materials, the heat accumulated in the BR slowing down the temperature relaxation in the structure. Figure 4(a) shows the pattern of the profile $\Delta T(x,t)$ under periodic heating of the structure with sinusoidal modulation of the heating laser power $I(t)$ with frequency 100 MHz, and the average flux density $\overline{I(t)} = 10^6$ W/cm² (which corresponds to the power of 1W focused on the area of 10 $\mu$m x 10 $\mu$m). There is a periodic modulation of $\Delta T$ with peak increase up to 80 °C near the interface between the metal and the BR, but the oscillation amplitude of $\Delta T$ for this case is about 40 °C: the temperature does not fall at the required rate. The variation of the position of the structure surface induced by thermal expansion reads as

$$\Delta x = \int \varepsilon \Delta T(x)dx. \tag{9}$$

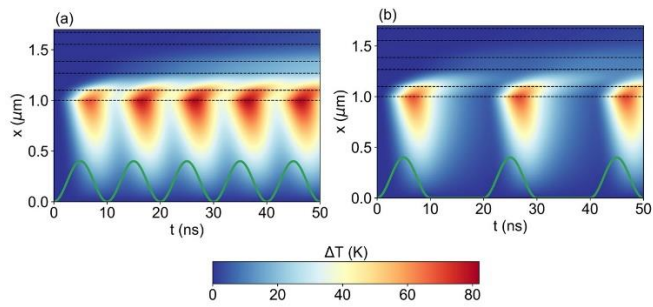

**Fig. 4.** Temporal and spatial distribution of the temperature increase $\Delta T(x,t)$ under periodic heating of the structure (a) with sinusoidal modulation of the laser power $I(t)$ with frequency 100 MHz, and average flux density $\overline{I(t)} = 10^6$ W/cm$^2$; (b) with a sequence of sinusoidal pulses with duration of 10 ns and repletion rate 50 MHz, and average flux density $\overline{I(t)} = 5 \cdot 10^5$ W/cm$^2$ Green line show the temporal dependence of laser flux density *I(t)*.

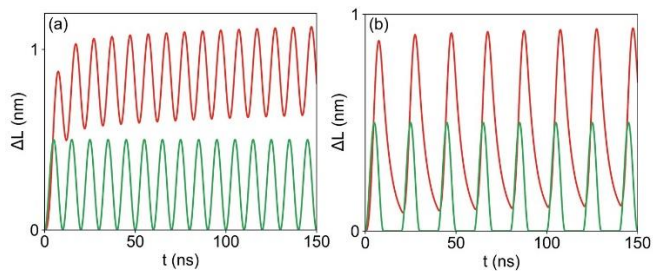

**Fig. 5.** Oscillation of the surface of the structure during heating by laser (a) with sinusoidal modulation of the laser power $I(t)$ with frequency 100 MHz, and average flux density $\overline{I(t)} = 10^6$ W/cm$^2$, (b) with a sequence of sinusoidal pulses with duration of 10 ns and repletion rate 50 MHz, and average flux density $\overline{I(t)} = 5 \cdot 10^5$ W/cm$^2$. Green line show the temporal dependence of laser flux density *I(t)*.

Figure 5a shows, that the shift of the surface by almost 1 nm at the first period of excitation followed by a sinusoidal oscillation with a magnitude of ~0.5 nm accompanied by a constant shift of the surface about 0.8 nm due to the heating of the structure as a whole. Note that for the parameters considered and the structure placed into a water-like medium the sound flux density is about 2 W/cm$^2$ in CW mode. Since the power of sound emitted by the oscillating surface, defined by equation 2, is proportional to the magnitude of oscillation squared, it would be advantageous to provide the cool-down of the structure to the initial level after the end of the heating pulse, which can be achieved by either the increase of the thermal conductivity of the metallic layer or by the reduction of the repetition rate. In Fig.4(b), the pattern of $\Delta T(x,t)$, i.e. the sequence of heating pulses with repetition rate of 50 MHz, is shown by green line. In this case the structure cools down to the initial level between pulses and the magnitude of the surface oscillation increases up to 0.9 nm, as shown in figure 5b.

In summary, the novel scheme of a photoacoustic generator based on the Tamm plasmon structure has been proposed. The Tamm plasmon structure provides a total absorption of the laser pulse at any desirable wavelength allowing the use of compact, powerful and feasible infrared semiconductor lasers with the wavelength about 1 μm for photoacoustic generators. Performance of various materials has been analyzed and it was shown that the structure utilizing magnesium for the active element possesses maximal efficiency. We have found that the efficiency of optical to sound power conversion grows linearly with flux density and modulation frequency of periodic excitation.

**Funding**. The work has been supported by "BRFFR-RFBR M-2019" № F19RM-006 "Study of 2D plasmonic nanostructures for photoacoustic transducers" RFBR grant 19-52-04005 "Two-dimensional plasmonic nanostructures for photoacoustic converter", the Swedish Research Council (Grant 2019-05154), Swedish Energy Agency (Grant 46563-1).

**Disclosures.** The authors declare no conflicts of interest.

28 Inhomogeneous and hogeneous heat equations (5) and (6) were solved nynumerically using "finite difference difference", [Finite Difference Methods in Heat Transfer, *M. Necati Özişik, Helcio R. B. Orlande, Marcelo J. Colaço, Renato M. Cotta,* 2nd Edition, Taylor and Francis Group. (2017)]
p129